\documentclass[reprint, superscriptaddress, amsmath,amssymb, aps, pra]{revtex4-1}

\usepackage{color}

\usepackage{natbib}
\usepackage{amsmath}

\usepackage{graphicx}
\usepackage{dcolumn}
\usepackage{bm}
\usepackage{siunitx}
\usepackage{upgreek}
\newcommand{\Dx}{\mathcal{D}x\,}
\newcommand{\Dxtilde}{\mathcal{D}\tilde{x}\,}
\newcommand{\DxjDxjtilde}{\prod_{j=1}^{d}\mathcal{D}x_j\mathcal{D}\tilde{x}_j\; }

\begin{document}

\preprint{APS/123-QED}

\title{Path integrals and nonlinear optical tweezers}

\author{B. Suassuna}
\email{bruno.b.suassuna@gmail.com}
\affiliation{Departamento de F\'{i}sica, Pontif\'{i}cia Universidade Cat\'{o}lica do Rio de Janeiro,  22451-900 Rio de Janeiro, RJ, Brazil}
\author{B. Melo}
\email{brunomelo@aluno.puc-rio.br}
\affiliation{Departamento de F\'{i}sica, Pontif\'{i}cia Universidade Cat\'{o}lica do Rio de Janeiro,  22451-900 Rio de Janeiro, RJ, Brazil}
\author{T. Guerreiro}
\email{barbosa@puc-rio.br}
\affiliation{Departamento de F\'{i}sica, Pontif\'{i}cia Universidade Cat\'{o}lica do Rio de Janeiro,  22451-900 Rio de Janeiro, RJ, Brazil}

\date{\today}


\begin{abstract}
We use path integrals to calculate perturbative corrections to the correlation function of a particle under the action of nonlinear optical tweezers, both in the overdamped and underdamped regimes. In both cases, it is found that to leading order nonlinearities manifest as shifts in the characteristic frequency of the system. The results are compared to numerical simulations. The present calculations enable a direct experimental method to access the nonlinear optical trap parameters by analyzing position data, similarly to standard harmonic tweezers.
\end{abstract}

\maketitle


\section{Introduction}

Optical tweezers are a widely used tool with applications in fundamental physics \cite{Moore2014, Monteiro2020a, Ether2015, Arvanitaki2013, Geraci2010, Moore2020}, chemistry \cite{Zhang2020, Cheuk2020} and biology \cite{Fazal2011, Nussenzveig2017, S.Araujo2019,Pontes2013, Armstrong2020}. The standard technique consists in using a tightly focused Gaussian laser beam to trap a dielectric particle immersed in some medium. To leading order approximation, the Gaussian profile creates a harmonic potential that confines the particle, which in turn undergoes Brownian motion due to interaction with its surroundings. Measuring the particle's position correlation function and its power spectrum, it is then possible to obtain information on the optical trap, notably its spring constant. This is commonly used as a calibration method essential to force-microscopy experiments \cite{Gieseler2020}.

Trapping the particle with beams that have complex intensity profiles is also an interesting possibility, and various optical potentials have been studied in the literature, such as the double-well landscape \cite{Rondin2017, Ciampini2020}, structured light beams with pattern revivals \cite{Silva2020}, bottle beams \cite{Melo2020}, frozen waves \cite{Suarez2020} and cylindrical vector beams \cite{Moradi2019}. The optical potentials generated by these \textit{structured light optical traps} are generally not harmonic, and the Brownian particle is subject to nonlinear force terms. Hence we shall refer to this type of trap as nonlinear optical tweezers. 

In the presence of nonlinearities, understanding the correlation functions and power spectrum of a trapped particle is not straightforward, as it involves nonlinear stochastic differential equations \cite{Dykman1984}. In this work, we address precisely this point: how do nonlinearities affect the particle's position correlation functions? 
To tackle this problem, we use perturbation theory methods and in particular the path integral formulation of stochastic differential equations \cite{Chow2015}. We study conservative non-linear forces, although non-conservative forces may also appear when dealing with large particles, due to scattering of the trapping beam. Such non-conservative forces are known to produce nonequilibrium steady states \cite{Amarouchene2019, Mangeat2019}.

After a brief introduction to the path integral formulation of Langevin equations, we apply the method to a number of different examples in one, two and three dimensions in the overdamped regime, later extending it to the underdamped regime.
We calculate corrections to the position correlation functions due to the presence of nonlinearities and show that in manifold situations the position power spectral density (PSD) of the trapped particle can be approximated to leading order by a Lorentzian function with a corner frequency that depends on the trap's parameters.
We provide explicit forms of this dependence for symmetric potential landscapes admitting a Taylor expansion, which can be used to both witness the presence of nonlinearities in an optical tweezer and calibrate the trap from experimental data. 
This Lorentzian approximation is of importance to precision tweezer experiments since, as we will show, fitting a Lorentzian function to PSD data while assuming a linear trap can lead to errors in trap stiffness calibration. 
We compare the results from perturbation theory to numerical simulations of different landscapes of interest. A discussion of future lines of investigation follows. 





\section{Path-integrals and Langevin equations}
The motion of a particle under the influence of a force field $\vec{F}(\vec{r})$  in a viscous medium
is governed by the Langevin equation
\begin{equation}
    m\ddot{\vec{r}}(t)=-\gamma \dot{\vec{r}}(t)+\vec{F}(\vec{r}(t))+\sqrt{2\gamma k_BT}\vec{\eta}(t),
    \label{eq:eom_3d}
\end{equation}
where $T$ is the temperature, $\gamma$ is the drag coefficient, $m$ is the particle's mass and $k_B$ is the Boltzmann constant. The last term in the right-hand side (RHS) represents environmental fluctuations. These are modelled using a Gaussian, white and isotropic stochastic process $\vec{\eta}(t)=(\eta_x(t), \eta_y(t), \eta_z(t))$, with zero mean and no correlations among different directions, i.e. $\langle \eta_i(t) \eta_j(t') \rangle= \delta_{ij}\delta(t-t')$.

For a sufficiently small particle, the inertial term is negligible in comparison to the drag term \cite{Jones2015}, and the system enters the so-called overdamped regime. In this case, Eq. \eqref{eq:eom_3d} can be approximated as

\begin{equation}
\label{eq:SDE3D}
    \dot{\vec{r}}(t) = \vec{f}(\vec{r}(t)) + \sqrt{D}\vec{\eta}(t),
\end{equation}
where $\vec{f}(x)=\vec{F}(x)/\gamma$ is the \textit{rescaled force} and $D=2 k_B T/\gamma$. When $\vec{f}(x)=(f_x(x), f_y(y), f_z(z))$, the motion along each direction is described by an independent equation of the form
\begin{equation}
    \dot{x}(t) = f(x(t)) + \sqrt{D}\eta(t),
    \label{eq:SDE}
\end{equation}
 which admits a path-integral formulation \cite{Chow2015,Wio2013}. For this equation, arbitrary moments $ \langle x(t_1)x(t_2)\ldots x(t_n) \rangle$ may be expressed as
\begin{equation}
  \int \Dx x(t_1)x(t_2)\ldots x(t_n) P[x],
\end{equation}
where $P[x]$ is a \textit{probability density functional} \cite{RichardP.Feynman2010} and possible paths of the particle are considered over some time interval $[-T,T]$, with initial condition $x(-T)=0$. The functional $P[x]$ may be represented by the path-integral expression
\begin{equation}
\label{Def P[x]}
    P[x] = \frac{\int \Dxtilde e^{-S[x,\tilde{x}]}}{\int \Dx\Dxtilde e^{-S[x,\tilde{x}]}},
\end{equation}
with the action functional $S[x,\tilde{x}]$ given by
\begin{equation}
    S[x,\tilde{x}] = \int_{-T}^{+T} \tilde{x}(t)\dot{x}(t) - \tilde{x}(t)f(x(t)) - \frac{D}{2}\tilde{x}(t)^2 dt.
\end{equation}
Eq. $\eqref{Def P[x]}$ should be understood as a formal limit of the corresponding expression in the discretized problem, defined accordingly to It\^o's prescription \cite{Chow2015,Gardiner2004}. In the discrete case, we have a random time series $x^{(N)}=(x_1,\ldots,x_N)$ with $x_k = x(-T+kh)$ and $h=2T/N$. The probability density associated with $x^{(N)}$ is
\begin{equation}
\label{Discrete Path Integral}
   \int \prod_{j=0}^{N-1}  \frac{dk_{j}}{2\pi} e^{-\sum_{j} ik_{j} \left[ \dfrac{x_{j+1} - x_{j}}{h}  - f_{j}  \right]  h  + \sum_{j} \frac{D}{2} (ik_{j})^{2} h},
\end{equation}
which goes into the path-integral equation in the formal limit $h\to 0$, where we let $ik_j \to \tilde{x}$ and $\prod_{j=0}^{N-1}  \frac{dk_{j}}{2\pi}\to \Dxtilde$.

In the subsequent calculations, we let $T\to +\infty$; this amounts to forgetting the previously mentioned initial condition $x(-T)=0$. Henceforth all time integrals are taken from $-\infty$ to $+\infty$. We are interested in calculating a perturbative expression for the auto-correlation function
\begin{equation}
    \langle x(t)x(0)\rangle = \frac{\int \Dx \Dxtilde x(t)x(0) e^{-S[x,\tilde{x}]}}{\int \Dx\Dxtilde e^{-S[x,\tilde{x}]}},
\end{equation}
for the case that $f(x)= -ax - p(x)$, where $a>0$ and $p(x)$ is a polynomial perturbation.
For this, we define the \textit{free moments} $\langle \prod_{j=1}^{n} x(t_j)\prod_{k=1}^{m}\tilde{x}(s_k)\rangle_0$ by the path-integral expression
\begin{equation}
    \frac{\int \Dx \Dxtilde \prod_j x(t_j)\prod_k\tilde{x}(s_k) e^{-S_0[x,\tilde{x}]}}{\int \Dx\Dxtilde e^{-S_0[x,\tilde{x}]}},
\end{equation}
where the free action is given by
\begin{equation}
    S_0[x,\tilde{x}] = \int \tilde{x}(t)\left(\frac{d}{dt} + a\right) x(t)\, dt.
\end{equation}
The relation between moments associated to the Langevin equation with force term $f(x)$ and the free moments is obtained by writing the total action as
\begin{equation}
    S[x,\tilde{x}] = S_0[x,\tilde{x}] + \int \tilde{x}(t)p(x(t)) dt - \frac{D}{2}\int\tilde{x}^2 dt,
\end{equation}
and, therefore,
\begin{equation}
\label{Fundamental Equation for the perturbative expansion}
  \small{  \langle x(t)x(0)\rangle = }\frac{\langle x(t)x(0) e^{-\int\tilde{x}(t')p(x(t'))dt'}e^{\frac{D}{2}\int\tilde{x}(t')^2 dt'}\rangle_0}{\langle e^{-\int\tilde{x}(t')p(x(t'))dt'}e^{\frac{D}{2}\int\tilde{x}(t')^2 dt'}\rangle_0}.
\end{equation}
By expanding the exponentials in the numerator as a power series we may compute approximations to the auto-correlation function through a perturbative series, with correction terms given by time integrals of the free moments. The denominator in Eq. \eqref{Fundamental Equation for the perturbative expansion} turns out to be equal to one to all orders in perturbation theory (see the Appendix for a derivation), and thus we shall omit it in what follows.

The free moment can be calculated by differentiation of the generating functional
\begin{equation}
    Z_0[J,\tilde{J}] = \int \Dx\Dxtilde e^{-S_0[x,\tilde{x}] + \int \tilde{J}(t) x(t) + J(t)\tilde{x}(t)\; dt},
\end{equation}
according to the expression,
\begin{equation}
\label{Free Moments as Functional Derivatives}
    \small{\frac{1}{Z_0[0,0]}\frac{\delta^{n+m} Z_0[J,\tilde{J}]}{\prod_{j}\delta \tilde{J}(t_j)\prod_{k}\delta{J}(s_k)} = \langle \prod_{j=1}^{n} x(t_j)\prod_{k=1}^{m}\tilde{x}(s_k)\rangle_0}.
\end{equation}
The generating functional $Z_0[J,\tilde{J}]$ is a Gaussian path-integral given by \cite{Chow2015},
\begin{equation}
    Z_0[J,\tilde{J}] = Z_0[0,0] \exp\Big(\int \tilde{J}(t)G(t,t')J(t')\, dt dt' \Big),
\end{equation}
where $G(t,t')=H(t-t')e^{-a(t-t')}$ and $H(t)$ is the left-continuous Heaviside function,
\begin{equation}
   H(t) = \begin{cases} 1 ,&t>0\\0,&t\leq 0\end{cases}\ \ .
\end{equation}
Therefore, the only non-vanishing free moments in Eq. $\eqref{Free Moments as Functional Derivatives}$ are those with $n=m$ and by Wick's theorem they are given by summing $\prod_i G(t_i,s_{j_i})$ over all possible pairings $\{(x(t_i),\tilde{x}(s_{j_i}))\}_{i=1,\ldots,n}$, where the indexes $\{j_1,\ldots,j_n\}$ are some permutation of $\{1,\ldots,n\}$. Note that any pairing that contains at least one equal time pair such as $(x(t'),\tilde{x}(t'))$ does not contribute to the computation since $G(t',t')=0$. As an example of Wick's theorem, the free moment $\langle x(t_1)x(t_2)\tilde{x}(s_1)\tilde{x}(s_2)\rangle_0$ is given by the expression
\begin{equation}
 G(t_1,s_1)G(t_2,s_2) + G(t_1,s_2)G(t_2,s_1).
\end{equation}

\section{One dimensional motion}

\subsection{Harmonic potential with quartic anharmonicity}
As a first example, we consider a Brownian particle in one dimension subject to a rescaled force of the form $f(x)= -ax - b_3 x^3$ and compute approximations to its auto-correlation function $\langle x(t)x(0)\rangle$. Using Eq. $\eqref{Fundamental Equation for the perturbative expansion}$, we find the leading order approximation,
\begin{equation}
\begin{split}
\langle x(t)x(0)\rangle \approx \frac{D}{2}\int dt_1 \langle \tilde{x}(t_1)^2x(t)x(0)\rangle_0 - \frac{D^2b_3}{8}\times \\ \int\, dt_1 dt_2 dt_3 \langle \tilde{x}(t_1)^2\tilde{x}(t_2)^2\tilde{x}(t_3)x(t_3)^3x(t)x(0)\rangle_0.\\
\,
\label{eq:first_order_ACF__linear_cubic}
\end{split}
\end{equation}

The first term in the RHS of Eq. \eqref{eq:first_order_ACF__linear_cubic} is equal to $(De^{-a|t|})/2a$, which is the  well-known result for the linear force $f(x)=-ax$. The second term corresponds to the leading order correction. To calculate the correction, we can draw diagrams organizing all possible pairings of variables generated by Wick's theorem. 


\begin{figure}
\centering
\includegraphics[width = \linewidth]{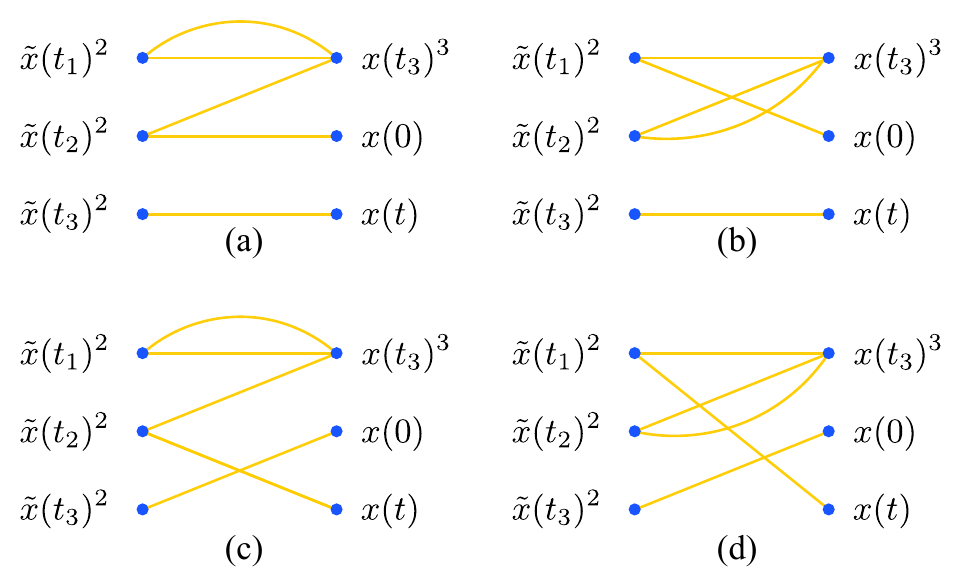}
    \caption{Diagrams corresponding to the leading order approximation in the case of a cubic nonlinearity. Each diagram corresponds to a term in Eq. \eqref{eq:4terms}.}
    \label{fig:Figure1}
\end{figure}

The diagrams are drawn by placing three points aligned in a left column representing the quantities $\{\tilde{x}(t_1)^2,\tilde{x}(t_2)^2,\tilde{x}(t_3)^2\}$ and three points in a right column representing $\{x(t_3)^3,x(0),x(t)\}$. Each quantity of the form $\tilde{x}(t_j)^m$ gives rise to $m$ edges, each of which must be connected to some point in the right. 
We draw all possible ways to connect the points in the left to the points in the right, omitting diagrams that connect $\tilde{x}(t_3)$ with $x(t_3)$. Each diagram is typically associated to multiple pairings, which contribute equally in the expansion by Wick's theorem. The multiplicities of each single diagram must be added into the final result as numerical pre-factors. The diagrams for the present example are shown in Fig. \ref{fig:Figure1}, and they correspond to the expansion
\begin{equation}
    \begin{split}
       \underbrace{3!2\langle\tilde{x}(t_1)x(t_3)\rangle_0^2\langle\tilde{x}(t_2)x(t_3)\rangle_0\langle\tilde{x}(t_2)x(t)\rangle_0\langle\tilde{x}(t_3)x(0)\rangle_0}_{(a)}\\  + \underbrace{3!2\langle\tilde{x}(t_1)x(t_3)\rangle_0^2\langle\tilde{x}(t_2)x(t_3)\rangle_0\langle\tilde{x}(t_2)x(0)\rangle_0\langle\tilde{x}(t_3)x(t)\rangle_0}_{(b)}\\  + \underbrace{3!2\langle\tilde{x}(t_1)x(t_3)\rangle_0\langle\tilde{x}(t_1)x(t)\rangle_0\langle\tilde{x}(t_2)x(t_3)\rangle_0^2\langle\tilde{x}(t_3)x(0)\rangle_0}_{(c)}\\ 
        + \underbrace{3!2\langle\tilde{x}(t_1)x(t_3)\rangle_0\langle\tilde{x}(t_1)x(0)\rangle_0\langle\tilde{x}(t_2)x(t_3)\rangle_0^2\langle\tilde{x}(t_3)x(t)\rangle_0}_{(d)},
    \end{split}
    \label{eq:4terms}
\end{equation}
from which we obtain the final form of the correlation function including corrections up to first order in $b_3$,
\begin{equation}
    \langle x(t)x(0)\rangle \approx \frac{D}{2a}e^{-a\vert t\vert} -\frac{3 b_3 D^2}{4a^3}e^{-a\vert t\vert}(1+a\vert t\vert)  
    .
\label{eq:corerlation1order}
\end{equation}

\begin{figure}[t]
    \centering
    \includegraphics[width  = \linewidth]{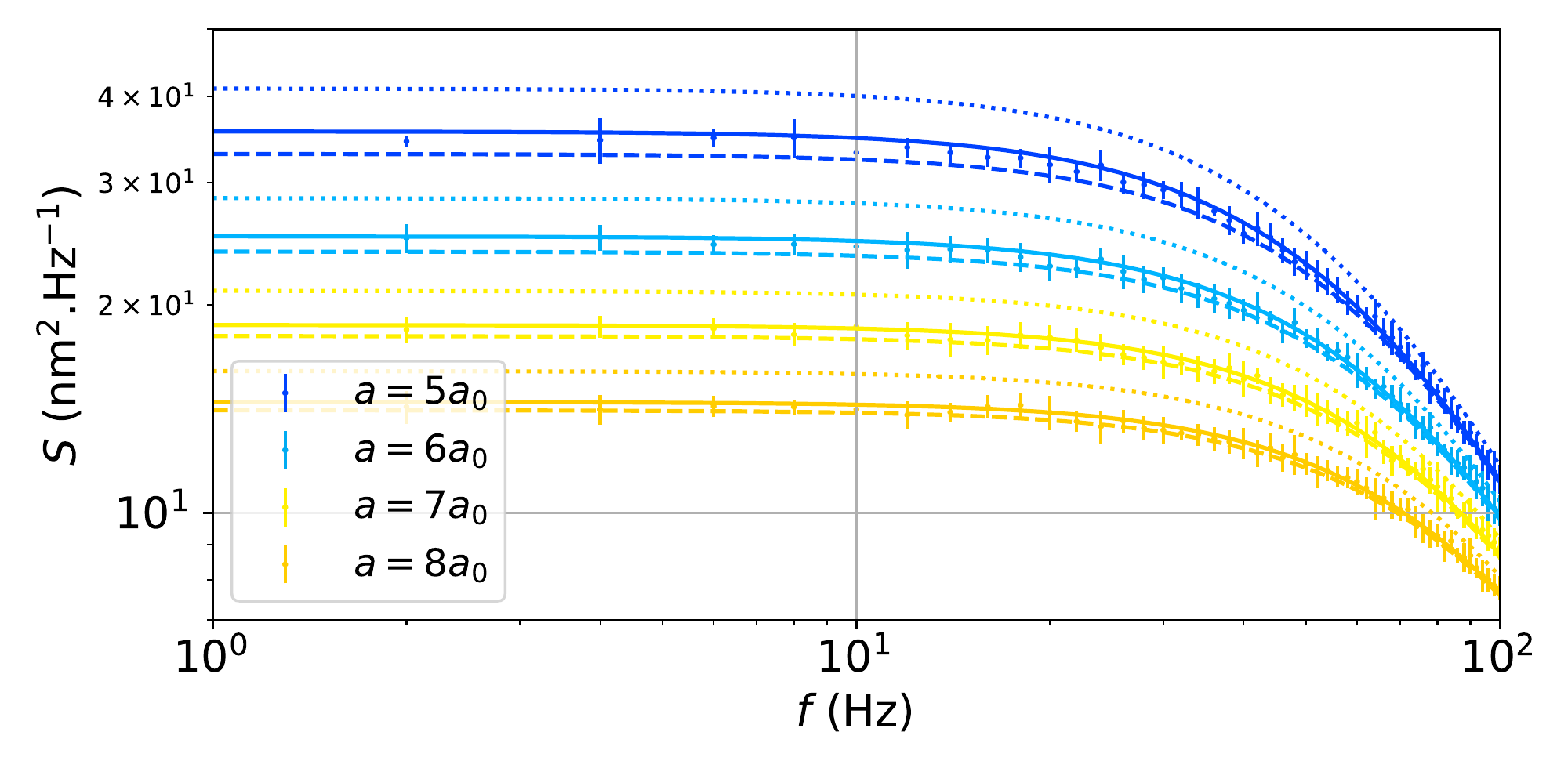}
    \caption{Comparison between the simulated PSD for $f(x)=-ax-b_3x^3$ and the theoretical expressions when no correction is added to the Lorentzian (dotted curves), when the leading order correction is added (dashed curves) and when corrections up to second order are added (solid curves). The values of $a$ and $b$ were varied while maintaining a fixed ratio between them.}
    \label{fig:Figure2}
\end{figure}

Fourier transforming the first term of the RHS of Eq. \eqref{eq:corerlation1order} yields the PSD 
with no perturbative correction
\begin{equation}
        S_0(\omega) =  \frac{D}{a^2+\omega^2},
        \label{eq:no_correction_linear_cubic}
    \end{equation}
and Fourier transforming the second term yields the PSD to first order in $ b_{3} $,
\begin{equation}
        S_1(\omega) = -\frac{3 b_3 D^2}{(a^2+\omega^2)^2}.
        \label{eq:first_correction_linear_cubic}
    \end{equation}
Going one order higher we get the quadratic correction in $ b_{3}$,
\begin{equation}
 \label{eq:second_correction_linear_cubic}
    S_2(\omega)=\frac{9 b_3^2 D^3}{(a^2+\omega^2)^3}\frac{59a^4+26a^2\omega^2-\omega^4}{4a^2(9 a^2+\omega^2)}.
\end{equation}

The different order approximations are displayed in Fig. \ref{fig:Figure2} together with PSD's calculated from simulated data. The temperature used in the plots and in the simulations was $295$K, while the drag coefficient was $1.26$fN.s/m, obtained by considering a sphere of radius $70$nm immersed in water, whose viscosity was calculated according to \cite{Fogelson2001}. Furthermore, four different values of $a$ were used: $4a_0$, $5a_0$, $6a_0$ and $7a_0$, where $a_0=79.4$s$^{-1}$, and the values of $b$ were chosen such that $b/a=4\times10^{12}$. Details regarding the simulations are presented in the Appendix.

In Fig. \ref{fig:Figure2}, the effect of the perturbative corrections becomes evident. If no correction is added, we see that the theoretical curve falls above the simulated data; if the leading order correction is considered, the theoretical curves move down, and fall slightly below the simulated data; when corrections up to second order are taken into account, the curves move up once again, falling closer to the simulation points. This is in agreement with the expectation that the predicted curve gets closer to the measured one as more perturbative corrections are contemplated.


To better evaluate  how many perturbative terms one has to consider to obtain a reasonable approximation, we can look at the changes on the PSD imposed by the first and second order corrections. From Eqs. \eqref{eq:no_correction_linear_cubic}, \eqref{eq:first_correction_linear_cubic} and \eqref{eq:second_correction_linear_cubic} we see that $S_{1}(\omega)/S_0(\omega)\to0$ and $S_{2}(\omega)/S_0(\omega)\to0$ when $\omega\gg a$. Therefore, the main changes in the PSD occur for $\omega< a$. Hence, one can look at the values of $S_0(0)$, $S_1(0)$ and $S_2(0)$ to decide if the first order approximation is sufficient or not. We have,
\begin{equation}
\begin{split}
        r_{1,0}&=\frac{S_1(0)}{S_0(0)}=-\frac{3b_3D}{a^2}\\
         r_{2,0}&=\frac{S_2(0)}{S_0(0)}=\frac{59b_3^2D^2}{4a^4}.
\end{split}
\end{equation}

Comparing the relative changes $r_{1,0}$ and $r_{2,0}$ with the precision of a specific experiment, one can conclude if the second order correction is appreciable or not.


Perturbation theory can be used to obtain higher order approximations to the PSD. For most applications, given the precision of typical experimental calibrations (on the order of a few \% \cite{BergSorensen2004}) the first order correction should be sufficient.



\subsection{Harmonic potential with higher order anharmonicities}

We may generalize the first order correction calculated in section III.A by considering the rescaled force,
\begin{eqnarray}
f(x) = -ax-b_{2k+1}x^{2k+1}\ .
\label{eq:general_force}
\end{eqnarray}
Force terms of this form can be generated by Taylor expanding a symmetric potential, for instance. The linear order correction to $\langle x(t)x(0)\rangle$ in the case of Eq. \eqref{eq:general_force} is obtained by integrating the expression
\begin{equation}
    \hspace{-0.5mm}\small{-\frac{b_{2k+1}D^{k+1}}{2^{k+1}(k+1)!}}\langle \tilde{x}(s)\tilde{x}(t_1)^2...\tilde{x}(t_{k+1})^2x(s)^{2k+1}x(t)x(0)\rangle_0,
\end{equation}
over $t_1$, ... , $t_{k+1}$ and $s$. Using Wick's theorem and performing the integrals we obtain (see the Appendix for details on the calculation):
\begin{equation}
\label{General linear order correction}
    -\frac{b_{2k+1}D^{k+1}}{2^{k}(k+1)!}\frac{(2k+2)!}{(2a)^{k+2}}e^{-a\vert t\vert}(1+a \vert t \vert).
\end{equation}
Therefore, the PSD including corrections up to first order in $b_{2k+1}$ is found to be,

\begin{multline}
     S(\omega) \approx \frac{D}{a^2+\omega^2} \\-\frac{b_{2k+1}D^{k+1}}{2^{2k}}\frac{(2k+2)!}{(k+1)!}\frac{1}{a^{k-1}}\frac{1}{(a^2+\omega^2)^2}.  
\end{multline}

\subsection{
General symmetric potentials}
The previous case can be further generalized by considering a potential with an arbitrary number of symmetric anharmonicities, which generates a rescaled force $f(x)=-ax-p(x)$ with 
\begin{equation}
    p(x) = \sum_{k=1}^{\ell} b_{2k+1}x^{2k+1}.
    \label{polynomial}
\end{equation}

To leading order in each constant $b_{2k+1}$, the perturbative correction to $\langle x(t)x(0)\rangle$ is the sum of each contribution given by Eq. \eqref{General linear order correction}, as shown in the Appendix. The leading order approximation to the PSD is then
\begin{equation}
\label{eq:Linear correction to PSD, polynomial perturbation}
    S(\omega)\approx \frac{D}{a^2+\omega^2} - 2a\delta\frac{D}{(\omega^2+a^2)^2},
\end{equation}
where
\begin{equation}
    \delta = \sum_{k=1}^{\ell} b_{2k+1}\left(\frac{D}{4a}\right)^k\frac{(2k+1)!}{k!}.
\end{equation}

Consider a particle trapped in a  potential that is symmetric around the origin, i.e. $V(x)=\phi(x^2)$, and suppose $\phi$ is analytic at $x=0$. The potential can be expanded as
\begin{equation}
    V(x)=\sum_{k=0}^\infty\frac{1}{k!}\phi^{(k)}(0)x^{2k},
\end{equation}
so we define $a=(2/\gamma) \phi^{(1)}(0)$ and, for all $k>0$, $b_{2k+1}=(2/\gamma) \phi^{(k+1)}(0)/ k!$. 
The PSD is then approximated by Eq. $\eqref{eq:Linear correction to PSD, polynomial perturbation}$, where 
\begin{equation}
   \delta=\frac{2}{\gamma}\sum_{k=1}^\infty \left(\frac{D}{4a}\right)^k\frac{(2k+1)!}{k!k!}\phi^{(k+1)}(0),
\end{equation}
provided the series converges.

Of special experimental interest is the Gaussian potential. In that particular case, $V(x)=V_0\exp{(-2x^2/\omega_0^2)}$ with $V_0<0$, which results in,
\begin{equation}
    \delta = a\sum_{k=1}^\infty\frac{(2k+1)!}{k!k!}\left(\frac{k_BT}{4V_0}\right)^k
\end{equation}
where $a=-4V_0/\gamma\omega_0^2$.

\subsection{The Lorentzian approximation}

Looking back at Eq. \eqref{eq:Linear correction to PSD, polynomial perturbation}, we notice that
\begin{equation}
    \frac{D}{\omega^2+(a+\delta)^2} = \frac{D}{\omega^2+a^2} - 2a\delta\frac{D}{(\omega^2+a^2)^2} + \mathcal{O}(\delta^2),
\end{equation}
from which we conclude that up to quadratic errors in $\delta$, the leading order approximation coincides with a Lorentzian distribution  of corner frequency $f_c=(a+\delta)/2\pi$. 
This shift on the corner frequency is analogous to the resonance frequency shift felt by an underdamped nanomechanical oscillator subjected to a Duffing nonlinearity \cite{Gieseler2013, Zheng2020}.

Qualitatively, a negative shift can be understood as an effective softening of the spring constant, which happens in the case in which the linear and nonlinear terms have opposite signs. On the other hand, a positive shift can be seen as an effective hardening of the spring constant.

 Moreover, the relative shift $\delta/a$ caused by a nonlinear term  $x^{2k+1}$ is proportional to $b_{2k+1}/a^{k+1}$. Therefore, if the constants $a$ and $b_{2k+1}$ are increased by the same factor - e.g., by using a larger trapping power - the relative shift decreases. This is expected, since the more confined the particle, the smaller its displacements are, making the role played by the nonlinearity less significant.



This dependence on trapping power suggests a method to identify the presence of nonlinearities in an optical trap using the Lorentzian approximation to the PSD. Taking into account the fact that $a$ and $b_{2k+1}$ are proportional to the trapping power $P$, the corner frequency becomes, 
\begin{equation}
    f_c(P) = \frac{a(P)+\delta(P)}{2\pi}= \alpha_0 P+\sum_{k=1}^l\frac{D^k}{P^{k-1}}\alpha_k,
\end{equation}
where $\alpha_k$ are constants. Hence, nonlinearities can be found by fitting Lorentzian curves to experimental data collected using different trapping powers. If the potential is quadratic, a simple linear dependence $f_c(P)=\alpha_0 P$ is expected. However, if this linear expression is insufficient to describe $f_c(P)$, the tweezer necessarily has nonlinear force terms. This provides a nonlinearity witness for an unknown potential.

Consider, for instance, the previously discussed quadratic potential perturbed by a quartic anharmonicity. The expected relation between the corner frequency and the trapping power is,
\begin{equation}
    f_c(P)=\alpha_0 P+\alpha_1,
    \label{eq:calibration}
\end{equation}
where $\alpha_1=3bD/4\pi a$. Measuring the corner frequency for different trapping powers and fitting it to Eq. \eqref{eq:calibration} allows one to obtain $a(P)$, by using $a=\alpha_0 P$. Additionally, one can obtain $b(P)$ through $b(P)=4\pi a(P)\alpha_1/3D$.

\begin{figure}
    \centering
    \includegraphics[width=\linewidth]{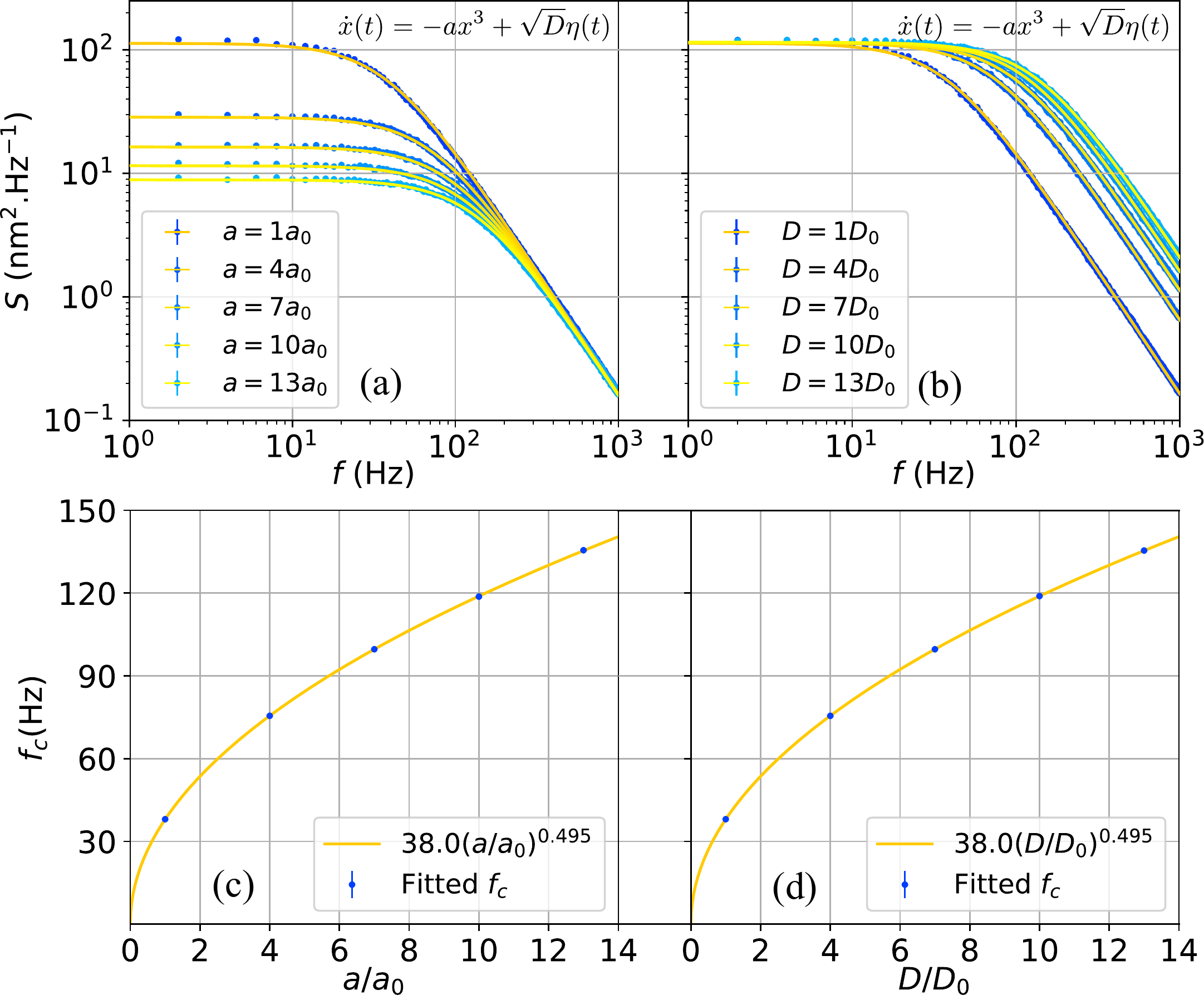}
    \caption{Simulation of the Brownian motion of a particle subject to the SDE $\dot{x}(t)=-ax^3+\sqrt{D}\eta(t)$. In part (a), the parameter $a$ is varied and $D=D_0$ is kept constant, while in part (b), $D$ is varied and $a=a_0$ is constant. Blue points correspond to the simulated data, while the yellow curves are Lorentzian fits. The corner frequencies obtained in the fits are displayed as blue points in parts (c) and (d). These corner frequencies were fitted to the functions $Ka^L$ in part (c), yielding $K=38.0$ and $L=0.495$, and $MD^N$ in part (d), yielding $M=38.0$ and $N=0.495$, showing a dependence approximately of the form $f_c\propto\sqrt{aD}$. 
    }
    \label{fig:Figure3}
\end{figure}

Using numerical simulations of the motion of a trapped particle, we find that the Lorentzian approximation is valid even when the nonlinear terms are large in comparison to the quadratic (linear-force) term, although the explicit form of $f_c(P)$ changes considerably. Fig. \ref{fig:Figure3} shows the PSDs calculated from simulations of the motion of a particle subject to the SDE,
\begin{equation}
    \dot{x}(t)=-ax^3+\sqrt{D}\eta(t)
\end{equation}
for different values of $a$ with a fixed $D$ (top) and distinct values of $D$ with fixed $a$ (bottom). The yellow curves are Lorentzian fits, which reveal a dependence approximately of the form $f_c\propto\sqrt{D}\sqrt{a}$. This dependence could be used, for instance, to calibrate purely cubic forces, such as the ones generated in bottle beams optical traps \cite{Melo2020}. The parameters used in this simulation were the same as the ones used for Fig. \ref{fig:Figure2}, except for $a_0=7.94\times10^{15}$m$^2$s$^{-1}$ and $D_0=6.47\times10^{-12}$m$^2$s$^{-1}$.

\section{Motion in higher dimensions}
The theory presented in the previous sections can be generalized to a particle moving in $d-$dimensions. We are primarily interested in the cases $d=2$ and $d=3$, but allowing for higher dimensions is useful when applying the method to systems with multiple particles, since a system of $N$ particles in $3$ dimensions is formally equivalent to a single particle in $d=3N$ dimensions. 

We denote the cartesian coordinates of the particle by $x = (x_1, \ldots, x_d)$ and assume they satisfy the SDE,
\begin{equation}
    \dot{x} = f(x) + \sqrt{D}\eta(t),
\end{equation}
where $f=(f_1,\ldots,f_d)$ is a deterministic force term and $\eta=(\eta_1,\ldots,\eta_d)$ is the isotropic white noise term.

Analogous to the one-dimensional case, correlation functions are given by the path integral expression,
\begin{equation}
    \langle x_k(t) x_\ell(0)\rangle = \frac{\int \DxjDxjtilde e^{-S[x,\tilde x]}x_k(t)x_\ell(0)}{\int\DxjDxjtilde e^{-S[x,\tilde x]}},
\end{equation}
where we define the response variables $\tilde{x} = (\tilde{x}_1, \ldots, \tilde{x}_d)$ and the action
\begin{equation}
    S[x,\tilde{x}] = \sum_{j=1}^{d} \int \tilde{x}_j\dot{x}_j - \tilde{x}_j f_j(x) - \frac{D}{2}\tilde{x}_j^2\; dt.
\end{equation}
Perturbation theory proceeds in a similar fashion to the one-dimensional case. We consider
\begin{equation}
    f_j(x) = - a_j x_j - p_j(x),
\end{equation}
for polynomial perturbations $p_j(x)$. We decompose the action as a free term $S_0$, 
\begin{equation}
    S_0[x,\tilde x] = \sum_{j=1}^{d}\int \tilde{x}_j(t')\left(\frac{d}{dt}+a_j\right)x_j(t')\; dt'\ ,
\end{equation}
plus a perturbation,
\begin{equation}
    \sum_{j=1}^{d}\int \tilde{x}_j(t') p_j(x(t'))\; dt' - \sum_{j=1}^{d}\frac{D}{2}\int \tilde{x}_j(t')^2\; dt' .
\end{equation}

Introducing the \textit{free moment generating functional},
\begin{multline}
Z_0[J,\tilde{J}] =\int   \DxjDxjtilde \times \\
 e^{-S_0 + \sum_{j}\int [\tilde{J}_j(t') x_j(t') + J_j(t')\tilde{x}_j(t') ]dt'},
\end{multline}
we proceed to define the \textit{free moments} of variables $x_i$ and $\tilde{x}_j$ as the appropriate functional derivatives of $Z_0$. This Gaussian path-integral is solved, yielding
\begin{equation}
    \small{Z_0[J,\tilde J] = \prod_{i,j=1}^{d}\exp(\int\int J_i(t)G_{i,j}(t,t')\tilde{J}_j(t')dtdt')},
\end{equation}
where 
\begin{equation}
G_{i,j}(t,t') = \delta_{ij} H(t-t') e^{-a_j(t-t')} = \langle x_i(t)\tilde{x}_j(t')\rangle_0 \ .    
\end{equation}

The perturbation theory is based on expansions of the expression,
\begin{equation}
  \langle x_k(t)x_\ell(0)\prod_{j=1}^{d} e^{-\int \tilde{x}_j(t')p_j(x(t'))dt'}e^{\frac{D}{2}\int \tilde{x}_j(t')^2 dt'}\rangle_0 \ ,
  \label{D-pert}
\end{equation}
and the resulting higher free moments can be evaluated using Wick's theorem, analogous to the one dimensional case. Notice that when considering all pairings between sets of variables $\{x_i(t_k)\}$ and $\{\tilde{x}_j(s_\ell)\}$, free moments of the form $\langle x_i(t)\tilde{x}_j(s)\rangle_0$ with $t=s$ or $i\neq j$ vanish due to the Heaviside function and Kronecker delta factors in $G_{ij}(t,s)$.
As in the one-dimensional case, Eq.\eqref{D-pert} should be divided by a normalization factor, analogous to the denominator in Eq. $\eqref{Fundamental Equation for the perturbative expansion}$. Such factor, however, is also equal to one to all orders in perturbation theory (as shown in the Appendix), and therefore we omit it. 

\begin{figure}[t!]
    \centering
    \includegraphics[width=\linewidth]{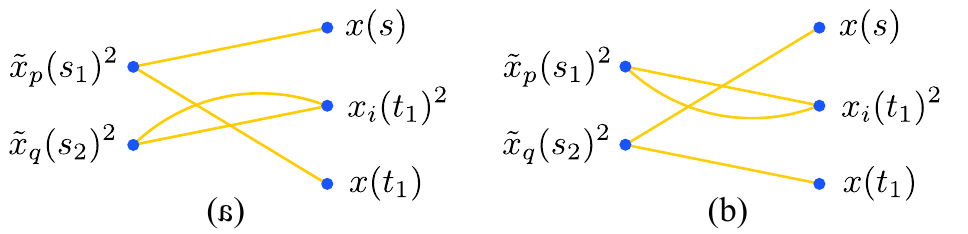}
    \caption{Diagrams corresponding to the expansion of the generic free moment $\langle x(s)\tilde{x}_p(s_1)^2\tilde{x}_q(s_2)^2 x_i(t_1)^2x(t_1)\rangle_0$, appearing when calculating corrections to the correlation functions in the d-dimensional case.}
    \label{fig:Figure4}
\end{figure}

\subsection{Radial harmonic potential with quartic anharmonicity}
As a first example in $ d = 2 $ and $3$ dimensions, we study the rotationally symmetric potential corresponding to the force
\begin{equation}
    f_j(x)= -ax_j - b\rho^2 x_j \ , 
\end{equation}
where,
\begin{eqnarray}
\rho =\left(\sum_{j=1}^{d} x_j^2\right)^{1/2}
\end{eqnarray}

The leading order correction in $b$ to the auto-correlator $\langle x_k(t) x_\ell(0)\rangle$ is given by
\begin{multline}
-\sum_{p,q,i,j}\frac{bD^2}{8} \int dt_{1} ds_1 ds_2 \times\\
\langle x_k(t)x_\ell(0)\tilde{x}_p(s_1)\tilde{x}_q(s_2)^2 x_i(t_1)^2\tilde{x}_j(t_1)x_j(t_1)\rangle_0 .
\end{multline}
As mentioned before, integrands for which $k\neq \ell$ vanish. Consider the case $k=\ell=1$; note that the only non-vanishing terms are the ones such that $j=1$. We rewrite the correction as
\begin{multline}
-\frac{bD^2}{8}\sum_{p,q,i=1}^{d}
    \int ds_1\, ds_2\, dt_1\,\times \\
    \langle x(t)x(0)\tilde{x}_p(s_1)^2\tilde{x}_q(s_2)^2 x_i(t_1)^2\tilde{x}(t_1)x(t_1)\rangle_0 \ .
\end{multline}

The terms with $i=1$ are non-vanishing only if $p=q=1$. Moreover the term with $i=p=q=1$ is identical to the leading order correction for the one-dimensional example presented in Section III.A. We now turn to the terms with $i\neq 1$. To expand those terms using Wick's theorem, note that $\tilde{x}(t_1)$ must be paired with $x(t)$ or $x(0)$. The total contribution of the pairings in which $\tilde{x}(t_1)$ is paired with $x(t)$ is $\langle x(0)\tilde{x}_p(s_1)^2\tilde{x}_q(s_2)^2 x_i(t_1)^2x(t_1)\rangle_0$ multiplied by $\langle x(t)\tilde{x}(t_1)\rangle_0$. The other pairings, in which $\tilde{x}(t_1)$ is paired with $x(0)$, give a total contribution of $\langle x(t)\tilde{x}_p(s_1)^2\tilde{x}_q(s_2)^2 x_i(t_1)^2x(t_1)\rangle_0$ multiplied by $\langle x(0)\tilde{x}(t_1)\rangle_0$. We expand the generic free moment $\langle x(s)\tilde{x}_p(s_1)^2\tilde{x}_q(s_2)^2 x_i(t_1)^2x(t_1)\rangle_0$ by two kinds of diagrams, which are shown in Fig. \ref{fig:Figure4}. The first vanishes unless $p=1$ and $q=i$ and the second vanishes unless $p=i$ and $q=1$; the non-vanishing diagrams contribute equally to the integral, so we may substitute 
\begin{equation}
\begin{split}
    \sum_{p,q}\sum_{i\neq 1} \langle x(s)\tilde{x}_p(s_1)^2\tilde{x}_q(s_2)^2 x_i(t_1)^2x(t_1)\rangle_0
\rightarrow \\
8\sum_{i\neq 1}\langle{x(s)\tilde{x}(s_1)}\rangle_0\langle{x(t_1)\tilde{x}(s_1)}\rangle_0\langle{x_i(t_1)\tilde{x}_i(s_2)}\rangle_0^2 \ .
\end{split}
\end{equation}

In conclusion, the leading order approximation to $\langle x(t)x(0)\rangle$ in $b$ is given by the result obtained in the one dimensional case corrected by the following terms:

\begin{multline}
    -bD^2\sum_{k\neq 1}\int dt_1\,ds_1\,ds_2\,  \langle\tilde{x}(t_1)x(t)\rangle_0\langle{x(0)\tilde{x}(s_1)}\rangle_0\\\times\langle{x(t_1)\tilde{x}(s_1)}\rangle_0\langle{x_k(t_1)\tilde{x}_k(s_2)}\rangle_0^2 \\
    -bD^2\sum_{k\neq 1}\int dt_1\,ds_1\,ds_2\, \langle\tilde{x}(t_1)x(0)\rangle_0\langle{x(t)\tilde{x}(s_1)}\rangle_0\\\times\langle{x(t_1)\tilde{x}(s_1)}\rangle_0\langle{x_k(t_1)\tilde{x}_k(s_2)}\rangle_0^2.
\end{multline}

Finally, $\langle x(t)x(0)\rangle$ is given by
\begin{equation}
    \langle x(t)x(0)\rangle = \frac{D}{2a}e^{-a\vert t\vert} -\frac{(d+2) b D^2}{4a^3}e^{-a\vert t\vert}(1+a\vert t\vert),
\end{equation}
leading to the PSD,
\begin{equation}
S(\omega)=\frac{D}{\omega^2+a^2}-\frac{bD^2(d+2)}{(\omega^2+a^2)^2},
\label{eq:S_rotationally}
\end{equation}
with $ d = 2 $ or $ 3 $.

\subsection{Anharmonic correction to a Gaussian trap}
Typical optical traps use a Gaussian beam to generate the trapping potential. Near the origin, this potential is approximately quadratic, but if the trapped particle moves sufficiently far from the center, higher-order terms become appreciable \cite{Gieseler2013, Zheng2020}. Perturbation theory in three-dimensions as outlined above can then be used to calculate corrections to the PSD of an overdamped particle in a Gaussian trap. We now discuss this type of correction. 

To fourth order in $x$, $y$ and $z$, the optical potential created by a Gaussian beam of wavelength $\lambda$ focused onto a particle of radius $R\ll\lambda$ is \cite{Jones2015},
\begin{equation}
    \frac{V(\rho,z)}{V_0}\approx1-\frac{2\rho^2}{\omega_0^2}-\frac{z^2}{z_R^2}+\frac{2\rho^4}{\omega_0^4}+\frac{z^4}{z_R^4}+\frac{4\rho^2z^2}{\omega_0^2z_R^2},
\end{equation}
where $\omega_0$ and $z_R$ are the the beams' waist and Rayleigh range, respectively, and
\begin{equation}
    V_0=-\frac{2\pi n_m R^3}{c}\frac{m^2-1}{m^2+1}I_0,
\end{equation}
with $I_0=2P/\pi\omega_0^2$ the optical intensity at the origin and $m=n_p/n_m$ the ratio between the refractive indices of the medium ($n_m$) and particle ($n_p$).

Motion of the particle is described by a SDE such as the one in Eq. \eqref{eq:SDE3D} with force components
\begin{equation}
\begin{split}
f_x(\vec{r})=-a_1x+b_1x^3+c_1xy^2+d_1xz^2\\
f_y(\vec{r})=-a_2y+b_2y^3+c_2yx^2+d_2yz^2\\
f_z(\vec{r})=-a_3z+b_3z^3+c_3zx^2+d_3zy^2,
\end{split}
\end{equation}
where $a_i=-4V_0/\gamma\omega_0^2$, $b_i=-8V_0/\gamma\omega_0^4$, $c_i=-8V_0/\gamma\omega_0^4$, $d_i=-8V_0/\gamma z_R^2\omega_0^2$ for $i=1,2$, and $a_3=-2V_0/\gamma z_R^2$, $b_3=-4V_0/\gamma z_R^4$, $c_3=-8V_0/\gamma z_R^2\omega_0^2$, $d_3=-8V_0/\gamma z_R^2\omega_0^2$. Note that since $V_0<0$ in a confining trap, all constants are positive.  

We compute $\langle x(t)x(0)\rangle$ to leading order in the constants $b_i,c_i,d_i$. As in the previous section, this is given by the result of the one-dimensional cubic perturbation in Eq. \eqref{eq:corerlation1order} plus corrections. The corrections are given by the integral of
\begin{multline}
   \frac{c_1D^2}{4}\langle x(t)x(0)x(t_1)\tilde{x}(t_1)\tilde{x}(s_1)^2\rangle_0\langle y(t_1)^2\tilde{y}(s_2)^2\rangle_0\; + \\ 
    \frac{d_1D^2}{4}\langle x(t)x(0)x(t_1)\tilde{x}(t_1)\tilde{x}(s_1)^2\rangle_0\langle z(t_1)^2\tilde{z}(s_2)^2\rangle_0 
\end{multline}
over $t_1,s_1,s_2$. After integration we find,
\begin{equation}
   \frac{c_1D^2 e^{-|t|a_1}(1+|t|a_1)}{4a_1^2a_2}+  \frac{d_1D^2 e^{-|t|a_1}(1+|t|a_1)}{4a_1^2a_3}.
\end{equation}

Analogous calculations for the $y$ and $z$ directions reveal the final expressions for the PSDs,
\begin{equation}
\begin{split}
S_x(\omega)\hspace{-1mm} =\hspace{-1mm} \frac{D}{a_1^2+\omega^2}\hspace{-1mm}+\hspace{-1mm}\left(\frac{3b_1}{a_1}\hspace{-1mm}+\hspace{-1mm}\frac{c_1}{a_2}\hspace{-1mm}+\hspace{-1mm}\frac{d_1}{a_3}\right)\hspace{-1mm} \frac{a_1D^2}{(a_1^2+\omega^2)^2}\\
S_y(\omega)\hspace{-1mm} =\hspace{-1mm} \frac{D}{a_2^2+\omega^2}\hspace{-1mm}+\hspace{-1mm}\left(\frac{3b_2}{a_2}\hspace{-1mm}+\hspace{-1mm}\frac{c_2}{a_1}\hspace{-1mm}+\hspace{-1mm}\frac{d_2}{a_3}\right)\hspace{-1mm} \frac{a_2D^2}{(a_2^2+\omega^2)^2}\\
S_z(\omega)\hspace{-1mm} =\hspace{-1mm} \frac{D}{a_3^2+\omega^2}\hspace{-1mm}+\hspace{-1mm}\left(\frac{3b_3}{a_3}\hspace{-1mm}+\hspace{-1mm}\frac{c_3}{a_1}\hspace{-1mm}+\hspace{-1mm}\frac{d_3}{a_2}\right)\hspace{-1mm} \frac{a_1D^2}{(a_3^2+\omega^2)^2}
\end{split}
\end{equation}

To see how these leading-order corrections affect measurements of the PSD, it is useful to consider the Lorentzian approximation discussed in Section II.D. Within the approximation, we have the following relative shifts in the corner frequencies,
\begin{equation}
\begin{split}
\frac{\delta_{x,y}}{a_{x,y}}=\frac{-6D/\omega_0^2}{-4V_0/\gamma\omega_0^2}=\frac{3D\gamma}{2V_0}=\frac{3k_BT}{V_0}\\
\frac{\delta_{z}}{a_{z}}=\frac{-5D/z_R^2}{-2V_0/\gamma z_R^2}=\frac{5D\gamma}{2V_0}=\frac{5k_BT}{V_0}.
\label{eq:relativeshifts}
\end{split}
\end{equation}

Since $k_BT\ll V_0$ is a necessary condition for stable trapping \cite{Li2013}, the relative shifts are usually small. For reasonable experimental parameters, however, they can become relevant. Using the values reported in \cite{Ashkin1986}, in which polystyrene spheres ($n_p=1.6$, $R=54.5$nm) are trapped in water ($n_m=1.33$) by an argon laser ($\lambda=514.5$nm, $12$mW) focused by an objective lens (NA$=1.25$), Eq. \eqref{eq:relativeshifts} predicts $\delta_x/a_x\approx-3.2\%$ and $\delta_z/a_z\approx-5.4\%$. For the silica spheres ($n_p=1.46$, $R=30$nm) trapped in the same work using a laser power of 200mW, the predicted shifts are $\delta_x/a_x\approx-2.3\%$ and $\delta_z/a_z\approx-3.9\%$. Therefore, had Lorentzian fits been applied to calibrate the trap stiffness, values between approximately $2\%$ and $5\%$ smaller than the real ones would have been found. Note that, since the fourth order and the sixth order terms of the potential have opposite sign, taking into consideration higher order terms would lead to relative shifts slightly smaller than the ones predicted in Eq. \eqref{eq:relativeshifts}.

Like in the one-dimensional case, the linear trap stiffness' $a_1$, $a_2$ and $a_3$ can be found by measuring and fitting the PSD's to Lorentzian functions at different trapping powers. The corner frequency of the PSD along the $i^{th}$ direction and the trapping power will follow a relation of the form $f_{c,i}(P)=\alpha_{0,i}P+\alpha_{1,i}$, where $\alpha_{0,i}=a_iP$. Therefore, fitting the corner frequencies to a linear function will yield a linear coefficient that can be used to calculate the linear trap stiffness for each value of $P$. 

\subsection{A structured beam for studying nonlinear effects}
\begin{figure}[t]
    \centering
    \includegraphics[width = \linewidth]{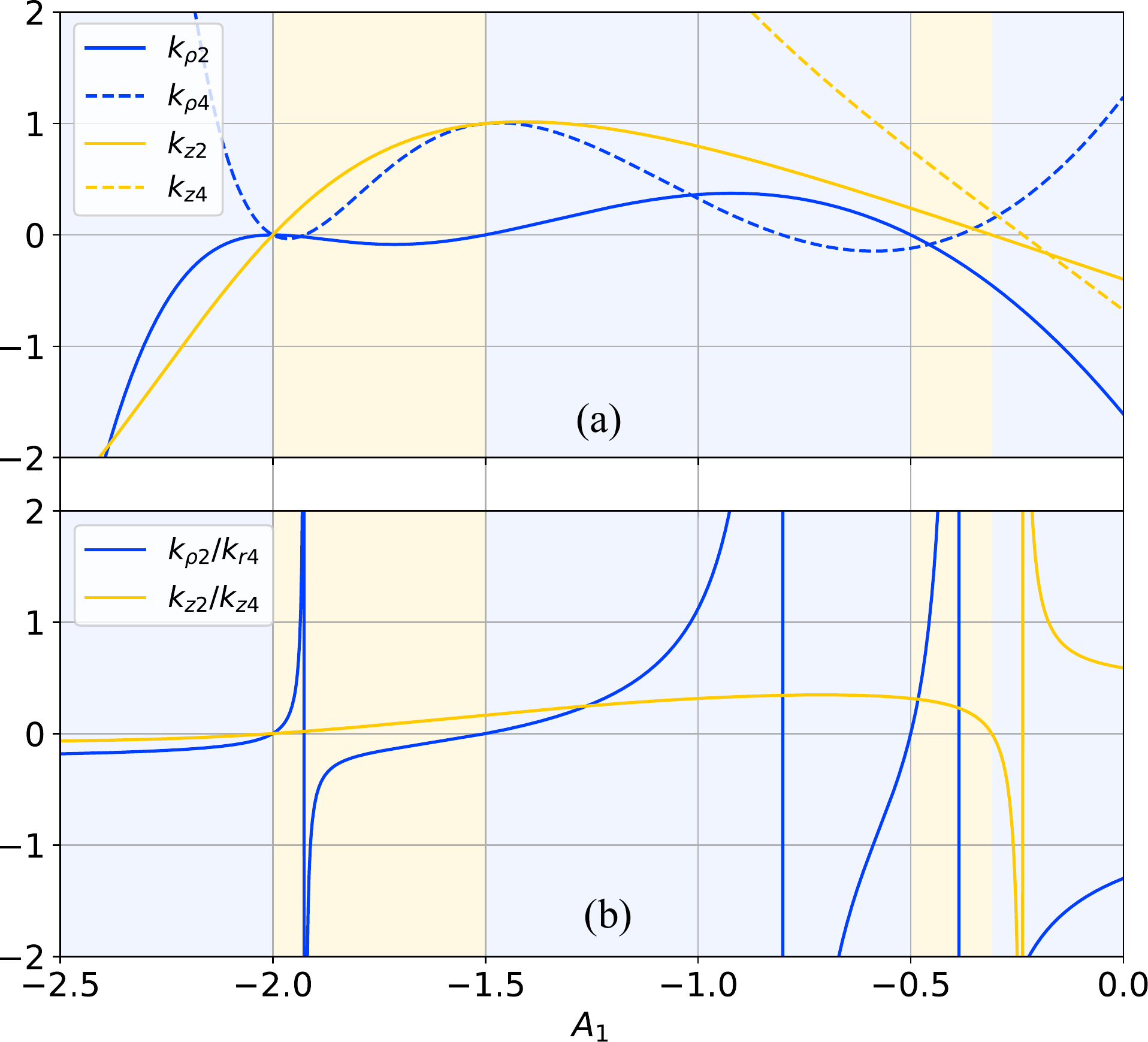}
    \caption{(a) Coefficients of the $\rho^2$, $\rho^4$, $z^2$ and $z^4$ terms found in the fourth-order expansion of the superposition of three Laguerre-Gauss beams: the first with $p=0$ and power $P_0$, the second with $p=1$ and power $A_1^2P_0$ and the third with $p=2$ and power $A_2^2P_0$. All beams have $\ell=0$ and $A_2$ is chosen such that there is no cross-term $\rho^2z^2$ in the expansion. (b) Ratios between the coefficients $k_{\rho2}$ and $k_{\rho4}$ and between $k_{z2}$ and $k_{z4}$.}
    \label{fig:Figure5}
\end{figure}

Although nonlinear effects manifest even in regular Gaussian traps, they can be enhanced by increasing the coefficients of the non-quadratic terms in the potential in comparison to the coefficients of the quadratic ones. 
This can be achieved by trapping particles with structured light beams, a technique which offers a panoply of potential landscapes \cite{Dholakia2008, Sun2008}. 

Consider for instance the intensity pattern generated by superposing three Laguerre-Gauss beams with $\ell=0$ and of equal polarization: the first with $p=0$ and power $P_0$, the second with $p=1$ and power $A_1^2P_0$ and the third with $p=2$ and power $A_2^2P_0$. By choosing $A_2=(-15-8A_1+\sqrt{220+220A_1+49A_1^2})/5$, the intensity around the origin can be approximated to fourth order as,
\begin{equation}
I(\rho,z) = I_0\hspace{-0.5mm} \left[k\hspace{-0.5mm}+\hspace{-0.5mm}k_{\rho2}\frac{\rho^2}{\omega_0^2}\hspace{-0.5mm}+\hspace{-0.5mm}k_{\rho4}\frac{\rho^4}{\omega_0^4}\hspace{-0.5mm}+\hspace{-0.5mm} k_{z2}\frac{z^2}{z_R^2}\hspace{-0.5mm}+\hspace{-0.5mm}k_{z4}\frac{z^4}{z_R^4}\right]
\end{equation}
where $\omega_0$ and $z_R$ are the beam's waist and Rayleigh range, respectively, $I_0=2P_0/\pi\omega_0^2$ and $k, k_{\rho2}, k_{\rho4}, k_{z2},k_{z4}$ are functions of $A_1$. Note that owing to the choice of $A_2$, the cross-term $\rho^2z^2$ has vanished.

In order for perturbation theory to apply the quadratic terms must dominate over of the quartic terms. Since this predominantly quadratic potential must be confining along the radial and axial directions, the factors $k_{\rho2}$ and $k_{z2}$ must have the same sign: if they are both positive, than the origin is a minimum of the intensity, and low refractive index particles ($ n_{p} < n_{m} $) can be trapped; if they are both negative, the origin is a maximum of the intensity, and high refractive index particles ($ n_{p} > n_{m} $) can be trapped. Fig. \ref{fig:Figure5}(a) shows $k_{\rho2}, k_{\rho4}, k_{z2}$ and $k_{z4}$ as functions of $A_1$, with blue regions highlighting when $k_{\rho2}$ and $k_{z2}$ have the same sign and yellow regions otherwise. 

Fig. \ref{fig:Figure5}(b) shows the ratios $k_{\rho2}/k_{\rho4}$ and $k_{z2}/k_{z4}$.
As it can be seen, as $A_1$ goes from $-1.5$ to approximately $-0.8$, the ratio $k_{\rho2}/k_{\rho4}$ goes from $0$ to $\infty$, while the ratio $k_{z2}/k_{z4}$ remains reasonably steady. Since all factors are positive in this interval, this region is ideal for trapping high refractive index particles, with the potential in the radial direction ranging from purely quartic to purely quadratic, and certainly encompassing a range of values of $A_1$ to which perturbation theory applies. Note also since there is no cross-term $\rho^2z^2$ in the potential, the forces generated by this superposition along the radial and axial directions would be decoupled, making this structured beam an ideal testbed for probing two dimensional Brownian motion and test nonlinear corrections such as the ones predicted by Eq. \eqref{eq:S_rotationally}.

\section{Beyond low Reynolds number}

In this section we generalize the path integral to the case in which the inertial term of the particle is non-negligible. In other words, we describe the motion of a Brownian particle immersed in a fluid of small viscosity (large Reynolds number) under the influence of a harmonic potential plus a nonlinear term under the approximations of perturbation theory. 

\subsection{A path integral for the second order equation}

We introduce a path-integral formulation for
\begin{equation}
\label{Second order Langevin Equation}
    \ddot{\vec{r}}(t)=-\kappa \dot{\vec{r}}(t)+\vec{f}(\vec{r}(t))+\sqrt{C}\vec{\eta}(t), 
\end{equation}
where $\kappa = \gamma/m$, $C=2\gamma k_B T/m^2$ and $\vec{f} = \vec{F}/m$ is the new rescaled force. 
For simplicity we shall work in one dimension, but we note that the generalization to higher dimensions is analogous to the one presented in Section IV. 

We reduce the order of the differential equation above by introducing the variable $v = \dot{x}$, satisfying
\begin{equation}
    \dot{v} = -\kappa v + f(x(t)) + \sqrt{C}\eta(t).
\end{equation}
As explained in the Appendix, we may write moments $ \langle x(t_1)x(t_2)\ldots x(t_n) \rangle$ as 
\begin{equation}
  \int \Dx x(t_1)x(t_2)\ldots x(t_n) P[x],
\end{equation}
where the probability density functional $P[x]$ is given by
\begin{equation}
\label{Path integral expression for P[x], second order case}
    P[x] = \frac{\int\mathcal{D}v P[x,v]}{\int\Dx\mathcal{D}v P[x,v]},
\end{equation}
with
\begin{equation}
\label{Path integral expression for P[x,v]}
    P[x,v] = \int\mathcal{D}\tilde{x}\mathcal{D}\tilde{v} e^{-S[x,v,\tilde{x},\tilde{v}]}.
\end{equation}
The action $S[x,v,\tilde{x},\tilde{v}]$ is defined as
\begin{equation}
\int \left(\tilde{x}(\dot{x}-v) + \tilde{v}\dot{v} + \kappa\tilde{v}v - \tilde{v}f(x) - \frac{C}{2}\tilde{v}^2\right) dt.
\end{equation}
 
Similarly to the previous sections, we assume that $f(x) = - \omega_0^2 x - \phi(x)$, for some small perturbation $\phi(x)$ with components $\phi_j(x)$ given by polynomials in $x$. In the \textit{underdamped} regime in which we are interested $\frac{\kappa^2}{4}<\omega_0^2$. Perturbation theory proceeds by defining the free action
\begin{equation}
S_0[x,v,\tilde{x},\tilde{v}] = \int \left(\tilde{x}(\dot{x}-v) + \tilde{v}\dot{v} + \kappa\tilde{v}v + \omega_0^2 \tilde{v}x\right) dt.
\end{equation}
The integrand above is of the form
\begin{equation}
    \begin{bmatrix}
    \tilde{x} && \tilde{v}
    \end{bmatrix} \left(I\frac{d}{dt} + A\right)  \begin{bmatrix}
    x \\ v
    \end{bmatrix},
\end{equation}
where the matrix $ A $ is given by
\begin{equation}
    A = \begin{bmatrix}
    0 && -1 \\
    \omega_0^2 && \kappa
    \end{bmatrix}.
\end{equation}
Therefore, the free moment generating functional $Z_0[J,\tilde{J},K,\tilde{K}]$, defined by
\begin{multline}
Z_0[J,\tilde{J},K,\tilde{K}] = \\
    \int\Dx\Dxtilde\mathcal{D}v\mathcal{D}\tilde{v}\; e^{-S_0 + J\cdot \tilde{x} + \tilde{J}\cdot x + K\cdot \tilde{v} + \tilde{K}\cdot v},
    \label{gauss-undergamped}
\end{multline}
is Gaussian-like, where the dot-product operation denotes $h_1\cdot h_2 = \int h_1(t)h_2(t)\, dt$, for functions $h_1(t)$ and $h_2(t)$.

The path-integral \eqref{gauss-undergamped} is solved by
\begin{equation}
  \mathcal{N}\exp\Big(\int  \begin{bmatrix}
    \tilde{J}(t) && \tilde{K}(t)
    \end{bmatrix} G(t,t')  \begin{bmatrix}
    J(t') \\ K(t')
    \end{bmatrix} \Big),
\end{equation}
where $\mathcal{N} = Z_0[0,0,0,0]$ and
\begin{equation}
    G(t,t') = H(t-t')e^{-A(t-t')}.
\end{equation}
We readily compute the matrix exponential and find,
\begin{multline}
   e^{-A\tau} = \\ e^{-\frac{\kappa\tau}{2}}\begin{bmatrix}
    \cos(\Omega\tau) + \frac{k\sin(\Omega\tau)}{2\Omega} && \frac{\sin(\Omega\tau)}{\Omega} \\
    -\omega_0^2\frac{\sin(\Omega\tau)}{\Omega} && \cos(\Omega\tau) - \frac{k\sin(\Omega\tau)}{2\Omega}
    \end{bmatrix},
\end{multline}
where $\Omega = \sqrt{\omega_0^2-\frac{\kappa^2}{4}}$. By Wick's theorem, free moments in the variables $x,v,\tilde{x},\tilde{v}$ are expanded in terms of the entries of the matrix $G(t,t')$, which are given by:
\begin{equation}
    G(t,t') = \begin{bmatrix}
    \langle x(t)\tilde{x}(t')\rangle_0 && \langle x(t)\tilde{v}(t')\rangle_0 \\
    \langle v(t)\tilde{x}(t')\rangle_0 && \langle v(t)\tilde{v}(t')\rangle_0
    \end{bmatrix}.
\end{equation}

Perturbation theory is developed by expanding arbitrary moments in terms of free moments. For auto-correlators, this is done by
\begin{equation}
\label{eq:Fundamental Equation for the perturbative expansion - second order case}
  \small{  \langle x(t)x(0)\rangle = }\frac{\langle x(t)x(0) e^{-\int\tilde{v}(t')\phi(x(t'))dt'}e^{\frac{C}{2}\int\tilde{v}(t')^2 dt'\rangle_0}}{\langle e^{-\int\tilde{v}(t')\phi(x(t'))dt'}e^{\frac{C}{2}\int\tilde{v}(t')^2 dt'}\rangle_0}.
\end{equation}
Note that this equation is remarkably similar to Eq. \eqref{Fundamental Equation for the perturbative expansion}, with the variable $\tilde{v}$ replacing the variable $\tilde{x}$ in the latter. Using Wick's theorem the free moments appearing in the expansion of $\langle x(t)x(0)\rangle$ are given in terms of
\begin{equation}
    g(t,t') = \langle x(t)\tilde{v}(t')\rangle_0 = e^{-\frac{\kappa(t-t')}{2}}\frac{\sin(\Omega(t-t'))}{\Omega}.
\end{equation}

The zero$^{\rm th}$ order approximation is evaluated as
\begin{equation}
    \langle x(t)x(0)\rangle = C \int g(t,s)g(0,s)\, ds.
    \label{eq:integral_zero_mass_case}
\end{equation}

For a harmonic potential, Eq. \eqref{eq:integral_zero_mass_case} results in,
\begin{equation}
    \langle x(t)x(0)\rangle\approx \frac{C e^{-\kappa\vert t\vert/2}(2 \Omega \cos{\Omega\vert t\vert}+\kappa \sin{\Omega\vert t\vert})}{\kappa\Omega(\kappa^2+4\Omega^2)},
\end{equation}
which gives the well-known PSD for the motion of a particle subject to Eq. $\eqref{Second order Langevin Equation}$ with $f(x)=-\omega_0^2 x$,
\begin{equation}
    S_0(\omega)= \frac{C}{\kappa^2\omega^2 + (\omega^2-\omega_0^2)^2}.
    \label{eq:zeroth_order_mas}
\end{equation}

\subsection{Quartic anharmonicty in a underdamped oscillator}

We can now investigate how a nonlinear rescaled force term $\phi(x)= - g x^3$ affects the PSD of an underdamped harmonic oscillator. According to Eq. \eqref{eq:Fundamental Equation for the perturbative expansion - second order case}, the leading order correction to the correlation function $\langle x(t)x(0)\rangle$ is given by the term,
\begin{multline}
    - \frac{C^2g}{8} \int dt_1 dt_2 dt_3\\\times \langle x(t)x(0)x(t_3)^3\tilde{v}(t_3)\tilde{v}(t_1)^2\tilde{v}(t_2)^2\rangle_0.
   \label{eq:first_order_ACF_mass}
\end{multline}

Solving the integral in Eq. \eqref{eq:first_order_ACF_mass} and Fourier transforming, we get the first order correction to the PSD,
\begin{equation}
    S_1(\omega)=\frac{3gC^2}{\kappa\omega_0^2}\frac{\omega^2-\omega_0^2}{(\kappa^2\omega^2 + (\omega^2-\omega_0^2)^2)^2}.
    \label{eq:first_order_mass}
\end{equation}

The expression for $S_0(\omega)+S_1(\omega)$ can be compared to the first order expansion,
\begin{multline}
    \frac{C}{\kappa^2\omega^2 + (\omega^2-(\omega_0+\delta)^2)^2}\approx\frac{C}{\kappa^2\omega^2 + (\omega^2-\omega_0^2)^2}\\
    +4C\omega_0\delta\frac{\omega^2-\omega_0^2}{(\kappa^2\omega^2 + (\omega^2-\omega_0^2)^2)^2},
    \label{eq:expansion_shifted_oscillator}
\end{multline}
which reveals that the effect of the anharmonicity, to first order, is to shift the resonance frequency of the harmonic oscillator by an amount,
\begin{equation}
    \delta = \frac{3gC}{4\kappa\omega_0^3}.
\end{equation}

Furthermore, the linear trap stiffness $k$ is related to the resonance frequency by $k=m\omega_0^2$. Therefore, if the trap stiffness is calculated by fitting the measured PSD to that of a linear oscillator, the resulting value will contain a shift approximately equal to,
\begin{equation}
    \delta_k=\frac{3gk_BT}{k}.
\end{equation}

Since both $g$ and $k$ are proportional to the trapping power, this shift is independent of $P$ in the regime in which the first order approximation suffices to describe the system. Fitting PSD's acquired using different values of $P$ will thus yield a linear dependence $k(P)=\alpha_0 P+\alpha_1$, with $\alpha_1\neq0$ if a quartic perturbation is part of the potential. Therefore, the nonlinearity witness derived for the overdamped regime also applies in the underdamped regime. Note that in practice the center frequency of the resonance peak varies in time around the shifted frequency $\omega_0+\delta$ \cite{Gieseler2013, Zheng2020}, and hence the value $\omega(P)$ used to calculate $k(P)$ should be understood as the average over many resonance frequencies measured using the same trapping power $P$.

\subsection{Underdamped oscillator in three dimensions}

Let us now consider an underdamped oscillator subject to the coupled set of equations,
\begin{equation}
\begin{split}
f_x(\vec{r})=-\omega_x^2x+g_1x^3+h_1xy^2+i_1xz^2\\
f_y(\vec{r})=-\omega_y^2y+g_2y^3+h_2yx^2+i_2yz^2\\
f_z(\vec{r})=-\omega_z^2z+g_3z^3+h_3zx^2+i_3zy^2.
\end{split}
\end{equation}

As in the overdamped case, the PSD of the motion along the $x$ axis is given by that of the linear case - which can be found using Eqs. \eqref{eq:zeroth_order_mas} and \eqref{eq:first_order_mass} -  plus the Fourier transform of the integral of,
\begin{multline}
   \frac{h_1C^2}{4}\langle x(t)x(0)x(t_1)\tilde{x}(t_1)\tilde{x}(s_1)^2\rangle_0\langle y(t_1)^2\tilde{y}(s_2)^2\rangle_0\; + \\ 
    \frac{i_1C^2}{4}\langle x(t)x(0)x(t_1)\tilde{x}(t_1)\tilde{x}(s_1)^2\rangle_0\langle z(t_1)^2\tilde{z}(s_2)^2\rangle_0.
\end{multline}

The final result is
\begin{multline}
    S_x(\omega) = \frac{C}{\kappa^2\omega^2 + (\omega^2-\omega_x^2)^2}\\-\left(\frac{3g_1C^2}{\kappa\omega_x^2}+\frac{h_1C^2}{\kappa\omega_y^2}+\frac{i_1C^2}{\kappa\omega_z^2}\right)\frac{\omega^2-\omega_x^2}{(\kappa^2\omega^2 + (\omega^2-\omega_x^2)^2)^2},
\end{multline}
which coincides, up to first order, with the PSD of a purely quadratic oscillator of frequency $\omega_x+\delta_x$, where
\begin{equation}
    \delta_x=-\frac{C}{4\kappa\omega_x}\left(\frac{3g_1}{\omega_x^2}+\frac{h_1}{\omega_y^2}+\frac{i_1}{\omega_z^2}\right).
\end{equation}

In a similar way, the PSDs of motion along the $y$ and $z$ directions coincide with that of an oscillator of frequency $\omega_y+\delta_y$ and $\omega_z+\delta_z$ respectively, where
\begin{equation}
    \begin{split}
       \delta_y&=-\frac{C}{4\kappa\omega_y}\left(\frac{3g_2}{\omega_y^2}+\frac{h_2}{\omega_x^2}+\frac{i_2}{\omega_z^2}\right)\\
       \delta_z&=-\frac{C}{4\kappa\omega_z}\left(\frac{3g_3}{\omega_z^2}+\frac{h_3}{\omega_x^2}+\frac{i_3}{\omega_y^2}\right).
    \end{split}
\end{equation}

\section{Discussion}

Knowledge of how nonlinear forces affect the position correlation functions of a Brownian particle enables a number of interesting applications. One can, for instance, envision precision force-microscopy experiments with a reduced systematic calibration error, aimed at measuring intrinsic fluctuations in biological phenomena \cite{Svoboda1994, Bialek2012}. 

\begin{figure}[t]
    \centering
    \includegraphics[width = 0.8\linewidth]{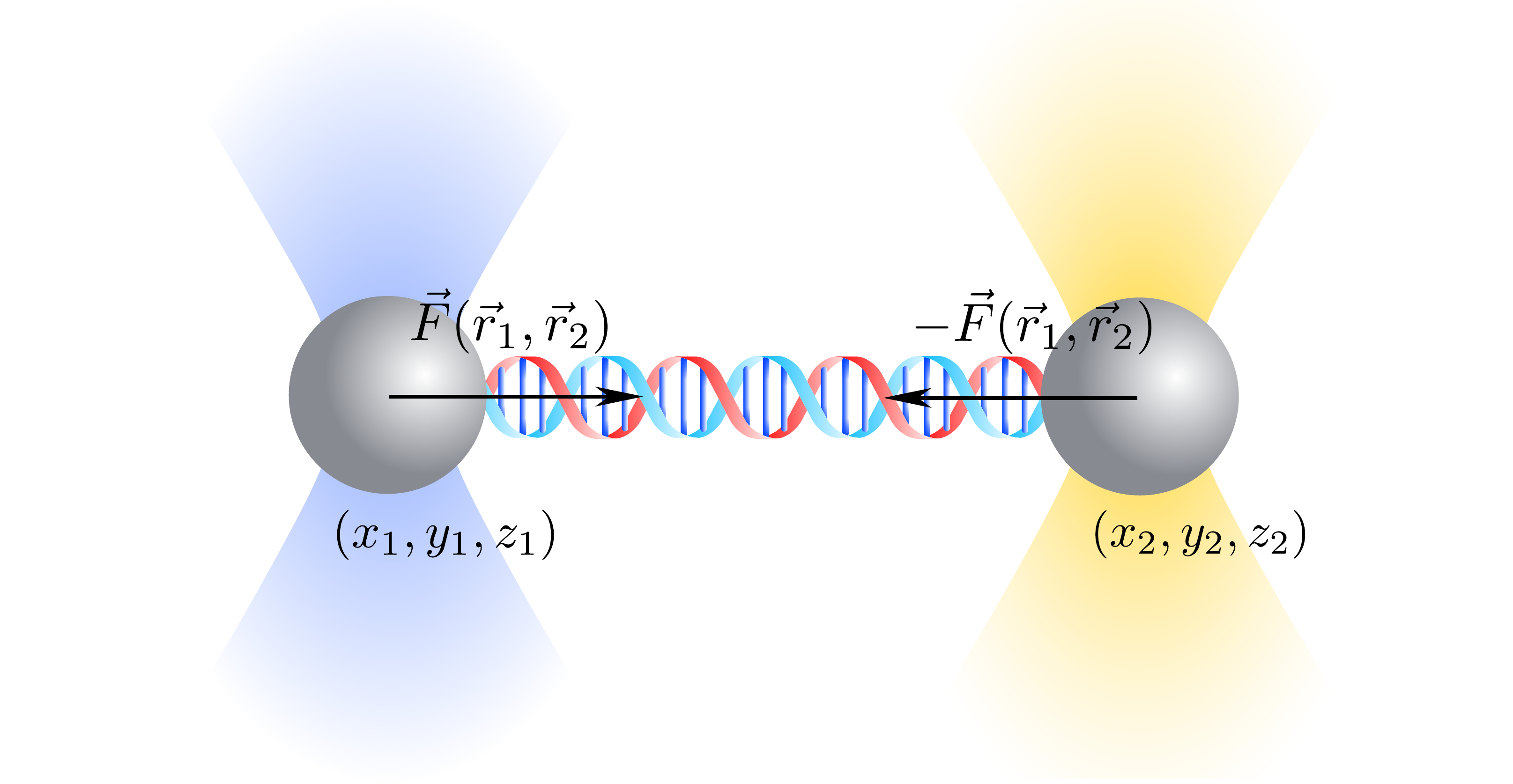}
    \caption{A nonlinearity witness may enable experiments aimed at characterizing nonlinear forces in optical tweezers through measurements of the particles' PSD, as in the case in which two beads are connected by a single DNA strand.}
    \label{fig:Figure6}
\end{figure}

The methods outlined in this work can also be used to obtain information on sources of nonlinear forces among Brownian particles. A prime example is that of two beads connected by a single strand of DNA \cite{Wang1997}, as outlined in Fig. \ref{fig:Figure6}. In this case, the DNA strand introduces a force of the form
\begin{eqnarray}
F = \left( \dfrac{k_{B}T}{L_{p}}   \right) \left[ \dfrac{1}{4(1 - x / L_{0})^{2}} - \dfrac{1}{4} + \dfrac{x}{L_{0}}   \right]
\end{eqnarray}
where $ L_{p}, L_{0} $ are parameters known as the persistence and countour lengths, respectively.  
In principle, by devising a nonlinearity witness as the one described in the main text, it is possible to obtain information on the force parameters by direct measurement of the particle's correlation functions and PSDs, without necessarily calibrating the traps' center and height \cite{Wang1997}.

The methods presented here can also find applications in the study of the so-called bottle beam optical trap: a structured light beam producing a dark focus, in which dielectric particles with a refractive index smaller than that of the surrounding medium can be trapped. Bottle beam traps have been suggested as a tool for trapping living cells \cite{Melo2020}. In that case, it would in principle be possible to measure the forces associated to cell division \textit{in vivo}. To properly measure such forces, one must understand the forces generated by the bottle trap, which turn out to be nonlinear. 

Nonlinearities are also expected to play a significant role in quantum optomechanics. An interesting future line of research is the effect of nonlinear potentials on ground state cooling. Deviations from a quadratic (linear-force) potential also become important when the particle is placed in controlled motion \cite{Pettit2019phononlaser, Gieseler2014}, and the effect of nonlinearities may be of relevance when preparing squeezed states of the mechanical oscillator through parametric pumping.

\section*{Acknowledgements}
We would like to acknowledge Lucianno Defaveri 
for enlightening discussions. 
This study was financed in part by the Coordena\c c\~{a}o de Aperfei\c coamento de Pessoal de N\'{i}vel Superior - Brasil (CAPES) - Finance Code 001, Conselho Nacional de Desenvolvimento Cient\'{i}fico e Tecnol\'{o}gico (CNPq), Funda\c c\~{a}o de Amparo \`a Pesquisa do Estado do Rio de Janeiro (Faperj, Scholarships No. E-26/202.830/2019), Instituto Serrapilheira (Serra-1709-21072).

\subsection*{Appendix A: Normalization constant}

We claim that, for any polynomial $p(x)$, the following equality holds in all orders of perturbation theory:
\begin{equation}
    \langle e^{-\int\tilde x p(x)} e^{\frac{D}{2}\int \tilde x^2}\rangle_0 = 1.
    \label{eq:normalization_constant_equals_1}
\end{equation}

Expanding the exponentials in the left hand side (LHS) of Eq. \eqref{eq:normalization_constant_equals_1} we obtain a perturbative series with integrals of all possible orders in $p$. The integrands of order $n$ are proportional to
\begin{equation}
   I_n=\langle \prod_{j=1}^{n}\tilde{x}(t_j)p(x(t_j))\prod_{k=1}^{m}\tilde{x}(s_k)^2\rangle_0, 
\end{equation}
which, for $n>0$, is a sum of terms proportional to
\begin{equation}
    J_{n,\sigma}=\prod_{j=1}^{n}\langle \tilde{x}(t_j)x(t_{\sigma(j))}\rangle_0,
\end{equation}
for some permutation $\sigma$ of the indexes $\{1,\ldots,n\}$. Now, $J_{n,\sigma}$ is non-vanishing only when
\begin{equation}
    t_1<t_{\sigma(1)},\,\, t_2<t_{\sigma(2)},\,\,  \ldots\,\,, t_n<t_{\sigma(n)},
    \label{eq:impossible_inequality}
\end{equation}
which we prove is impossible. Suppose the inequalities in Eq. \eqref{eq:impossible_inequality} hold for some permutation $\sigma$. Consider a permutation $\pi$ such that
\begin{equation}
    t_{\pi(1)}\leq t_{\pi(2)}\leq \ldots\leq t_{\pi(n)}.
\end{equation}

By assumption, $t_{\pi(n)}<t_{\sigma(\pi(n))}$. However, $\sigma(\pi(n))=\pi(\ell)$ for some index $\ell\in\{1,\ldots,n\}$, namely $\ell = \pi^{-1}(\sigma(\pi(n)))$. Therefore, we find that
\begin{equation}
    t_{\pi(n)} < t_{\pi(\ell)},
\end{equation}
which is a contradiction. Hence, $J_{n,\sigma}$ vanishes for $n>0$ and any permutation $\sigma$, which implies that $I_n=0$ for all $n>0$. We conclude that the only non-vanishing term in the expansion of the LHS of Eq.\eqref{eq:normalization_constant_equals_1} is the zero$^{\rm th}$ order term, which is equal to 1, proving our initial claim.

We extend this result for the multi-dimensional case. Suppose we have $\tilde{x},x,p(x)\in\mathbb{R}^d$, with each component of $p(x)$ a polynomial in the components $x_1,\ldots,x_d$ of $x$. We claim:
\begin{equation}
    \langle e^{-\sum_k\int\tilde x_k p_k(x)} e^{\frac{D}{2}\sum_k\int \tilde x_k^2}\rangle_0 = 1,
\end{equation}
in all orders of perturbation theory.

Expanding the exponentials, we get integrands proportional to
\begin{equation}
   I^{(d)}_n=\langle \prod_{j=1}^{n}\tilde{x}_{m_j}(t_j)p_{m_j}(x(t_j))\prod_{k=1}^{m}\tilde{x}_{n_k}(s_k)^2\rangle_0, 
\end{equation}
for some $m_1,\ldots,m_n,n_1,\ldots,n_m\in\{1,\ldots,d\}$. This is proportional to
\begin{equation}
    J^{(d)}_{n,\sigma}=\prod_{j=1}^{n}\langle \tilde{x}_{m_j}(t_j)x_{m_{\sigma(j)}}(t_{\sigma(j))}\rangle_0,
\end{equation}
which is zero when $n>0$, for any permutation $\sigma$, by the same argument used for the one-dimensional case.

\subsection*{Appendix B: Harmonic potential with a higher order anharmonicities}\label{First order correction}
Let us consider the SDE in Eq.(\ref{eq:SDE}) with
\begin{equation}
    f(x) = -ax-bx^{2k+1}.
\end{equation}
For this potential, the linear correction in $b$ is obtained by integrating the expression,
\begin{equation}
    -\frac{bD^{k+1}}{2^{k+1}(k+1)!}\langle \tilde{x}(s)\tilde{x}(t_1)^2...\tilde{x}(t_{k+1})^2x(s)^{2k+1}x(t)x(0)\rangle,
\end{equation}
 over the ``internal points'' $\{s,t_1,...,t_{k+1}\}$. In order to do so, we need to find all sets of pairs in which the first element of the pair belongs to the list $L_1=[s, t_1, t_1, t_2, t_2, ...,t_{k+1}, t_{k+1}]$ and the second one belongs to $L_2=[s,s,s,...,s,t,0]$.

First, note that if the element $s$ from $L_1$ is paired with an element $s$ from $L_2$, the set of pairs will have null contribution, due to the fact that $G(s,s)=0$. Therefore, the only contributing sets are those in which the element $s\in L_1$ is paired with either $t$ or $0$. Let us break the problem into these two classes of sets.

\subsubsection*{The element $s$ is paired with $t$}
There are $(2k+2)!$ sets of pairs containing the element $(s,t)$. This can be easily seen if we consider the ordered list formed by the remaining elements of $L_2$: $[s,s,..., s,0]$. There are $2k+2$ possible elements in $L_1$ to pair with the first element of the ordered list, $2k+1$ possible elements in $L_1$ to pair with the second element, and so on. Once all the pairs are chosen, we have the following situation:
\begin{itemize}
    \item $s$ has been paired with $t$;
    \item some element $t_j$ has been paired with $0$, while the other element $t_j$ has been paired with $s$;
    \item the other elements $t_{i\neq j}$ have been paired with $s$.
\end{itemize}
Therefore, the contribution of a single list of pairs is
\begin{multline}
    I_A(t) = \int_{-\infty}^{\infty}\,ds\,dt_1...dt_{k+1}\\G(s,t)G(t_j,0)G(t_j,s)\prod_{\substack{i=1\\i\neq j}}^{k+1}G(t_i,s)^2.
\end{multline}
The integrals over $t_{i\neq j}$ can be easily calculated:
\begin{equation}
    \int_{-\infty}^{\infty}dt_iG(t_i,s)^2=\frac{1}{2a}.
\end{equation}
Performing the integrals over all such $t_i$'s yields
\begin{equation}
\label{eq:almost there}
    I_A(t) =\left(\frac{1}{2a}\right )^k\int_{-\infty}^{\infty}ds\,dt_jG(s,t)G(t_j,0)G(t_j,s).
\end{equation}
Integrating over the remaining variables $s$ and $t_j$, we find,
\begin{equation}
    I_A(t)=\left(\frac{1}{2a}\right)^{k+2}\begin{cases}e^{-a t}, & t>0\\e^{a t}(1-2a t), &t<0\end{cases}.
\end{equation}
Recalling that there are $(2k+2)!$ lists of pairs, the total contribution of this class of lists is $(2k+2)!I_A(t)$.

\subsubsection*{The element $s$ is paired with $0$}
By an argument analogous as the one used in the previous case, the contribution of a generic list of pairs in the present one is
\begin{multline}
    I_B(t) = \int_{-\infty}^{\infty}\,ds\,dt_1...dt_{k+1}\\G(s,0)G(t_j,t)G(t_j,s)\prod_{\substack{i=1\\i\neq j}}^{k+1}G(t_i,s)^2, 
\end{multline}
which after integration over $t_{i\neq j}$ results in
\begin{equation}
     I_B(t)=\left(\frac{1}{2a}\right)^{k}\int_{-\infty}^{\infty}ds\,dt_jG(s,0)G(t_j,t)G(t_j,s),
\end{equation}

and after integration over the remaining variables,
\begin{equation}
    I_B(t)=\left(\frac{1}{2a}\right)^{k+2}\begin{cases}e^{-a t}(1+2a t), & t>0\\e^{a t}, &t<0\end{cases}.
\end{equation}

Hence, we have a total contribution of $(2k+2)!I_B(t)$ by this second class of sets of pairs.

Putting it all together, we find that the total first-order correction to the autocorrelation function is
\begin{equation}
    -\frac{bD^{k+1}}{2^{k}(k+1)!}\frac{(2k+2)!}{(2a)^{k+2}}e^{-a\vert t\vert}(1+a \vert t \vert)
\end{equation}
and, finally, the first order correction to the PSD is
\begin{equation}
   S^{(2k+1)}_1(\omega)= -\frac{bD^{k+1}}{2^{2k}}\frac{(2k+2)!}{(k+1)!}\frac{1}{a^{k-1}}\frac{1}{(a^2+\omega^2)^2}.
\end{equation}


\subsection*{Appendix C: General symmetric potential}

We show that a perturbation of the form 
\begin{equation}
    p(x) = \sum_{k=1}^{\ell} b_{2k+1}x^{2k+1},
\end{equation}
leads to a correction of the auto-correlator that is in leading order in the constants $\{b_{2k+1}\}$ simply the sum of the contribution of each monomial. By Eq. $\eqref{Fundamental Equation for the perturbative expansion}$, the auto-correlator is given by
\begin{multline}
  \langle x(t)x(0)\rangle =\\  \langle x(t)x(0) \prod_{k=1}^{\ell} e^{-b_{2k+1}\int\tilde{x}(t')x(t')^{2k+1}dt'}e^{\frac{D}{2}\int\tilde{x}(t')^2 dt' \rangle_0}.
\end{multline}
The claim follows once we notice
\begin{multline}
  \prod_{k=1}^{\ell} e^{-b_{2k+1}\int\tilde{x}(t')x(t')^{2k+1}dt'} =\\  1 - \sum_{k=1}^{\ell} b_{2k+1}\int\tilde{x}(t')x(t')^{2k+1}dt' + \ldots,
\end{multline}
where the remaining terms are all of higher order in the constants $\{b_{2k+1}\}$.
\subsection*{Appendix D: A structured beam for studying nonlinear effects}

The electric field of a Laguerre-Gauss (LG) beam is
\begin{eqnarray}
  \label{eq:LGbeam2}
      \hspace*{-2em}&\,&E^{LG}_{\ell,p}(\rho,\phi,z) =\sqrt{\frac{4P}{c\epsilon\pi\omega(z)^2}}\sqrt{\frac{p!}{(\vert \ell \vert+p)!}}\times\nonumber\\&&\left( \frac{\sqrt{2}\rho}{\omega(z)} \right)^{\vert\ell\vert}L^{\vert\ell\vert}_p\left( \frac{2\rho^2}{\omega(z)^2} \right)\exp\left[-\frac{\rho^2}{\omega(z)^2}\right]\times\nonumber\\&&\exp[ik_mz+ik_m\frac{\rho^2}{2R(z)}-i\zeta(z)+i\ell\phi],
\end{eqnarray}
where $c$ is the speed of light, $\epsilon$ is the medium's permittivity, $P$ is the beam's power, $k_m$ is the wavenumber in the medium and $\omega(z)$, $R(z)$, $\zeta(z)$ and $L^{\vert\ell\vert}_p$ are the beam width, the wavefront radius, the Gouy phase and the Associated Laguerre polynomial. These quantities are respectively given by
\begin{eqnarray}
\omega(z)&=&\omega_0\sqrt{1+\frac{z^2}{z^2_R}};\\
  R(z) &=& z\left(1+\frac{z_R^2}{z^2} \right);\\
\zeta(z)&=&(2p+\vert\ell\vert+1)\arctan \frac{z}{z_R};\\
    L_p^{\vert\ell\vert}(x) &=& \sum^p_{i=0}\frac{1}{i!}\binom{p+\vert\ell\vert}{p-i}(-x)^i
\end{eqnarray}
\noindent where $z_R$ is the the Rayleigh range and $\omega_0$ is the beam's waist.

Consider the superposition of three LG beams of $\ell=0$ and equal polarization: the first with $p=0$ and power $P_0$, the second with $p=1$ and power $A_1^2P_0$ and the third with $p=2$ and power $A_2^2P_0$, where $A_1, A_2\in\mathbb{R}$. Expanding the intensity of such beam to fourth order around the origin results in 
\begin{multline}
\frac{I(\rho,z)}{I_0} =\\\hspace{-0.5mm} k\hspace{-0.5mm}+\hspace{-0.5mm}k_{\rho2}\frac{\rho^2}{\omega_0^2}\hspace{-0.5mm}+\hspace{-0.5mm}k_{\rho4}\frac{\rho^4}{\omega_0^4}\hspace{-0.5mm}+\hspace{-0.5mm} k_{z2}\frac{z^2}{z_R^2}\hspace{-0.5mm}+\hspace{-0.5mm}k_{z4}\frac{z^4}{z_R^4}+k_{\rho2z2}\frac{\rho^2z^2}{\omega_0^2z_R^2},
\end{multline}
where $I_0=2P_0/\pi\omega_0^2$ and $k$, $k_{\rho2}$, $k_{z2}$, $k_{\rho4}$, $k_{z4}$, $k_{\rho2z2}$ are functions of $A_1$ and $A_2$. By choosing $A_2=(-15-8A_1+\sqrt{220+220A_1+49A_1^2})/5$, we get $k_{\rho2z2}=0$. Thus, if this superposition is used to trap a dielectric particle, the forces and the motions along the radial and axial directions will be decoupled, allowing one to probe two dimensional Brownian movement. 

Note that since $220+220A_1+49A_1^2>0$ for all $A_1\in \mathbb{R}$, the aforementioned relation between $A_1$ and $A_2$ implies that $A_2\in\mathbb{R}$ for all $A_1\in \mathbb{R}$, in conformity with our initial assumption.

\subsection*{Appendix E: A path integral for the second order equation}

As mentioned in the main text, we interpret the second order stochastic equation
\begin{equation}
    \ddot{x}(t)=-\kappa \dot{x}(t)-f(x(t))+\sqrt{C}\eta(t),
\end{equation}
as the system of first order equations
\begin{equation}
    \begin{cases}
    \dot{x} = v \\
    \dot{v} = -\kappa v + f(x(t)) + \sqrt{C}\eta(t)
    \end{cases}.
\end{equation}

The equations above are then discretized accordingly to Ito's prescription,
\begin{equation}
\label{Discrete version of second order SDE}
    \begin{cases}
    x_{n+1} - x_n = v_n h \\
    v_{n+1} - v_n = -\kappa v_n h + f_n h + \sqrt{C} w_n \sqrt{h}
    \end{cases},
\end{equation}
where we divide the interval $[-T,T]$ in $N$ subintervals of size $h=2T/N$, impose $x(-T)=v(-T)=0$, let
\begin{equation}
    x(-T+kh) = x_k \ \ , \ \ v(-T+kh) = v_k,
\end{equation}
$f_k=f(x_k)$ and suppose the increments $\{w_n\}_{n=0,\ldots,N-1}$ are independent Gaussian variables of mean $0$ and variance $1$. As mentioned in the main text, we later take the limit $T\to\infty$, which erases the imposed initial conditions.

Let $x^{(N)}=(x_1,\ldots,x_N)$, $v^{(N)}=(v_1,\ldots,v_N)$ and $w^{(N)}=(w_0,\ldots,w_{N-1})$. Conditioned to some values of $w^{(N)}$, the joint probability density associated with $x^{(N)}$ and $v^{(N)}$ is
\begin{multline}
    P(x^{(N)},v^{(N)}\vert w^{(N)}) = \prod_{n=0}^{N-1}\delta(x_{n+1}-x_n-v_n h)\times \\ \delta(v_{n+1} - v_n + \kappa v_n h - f_n h - \sqrt{C} w_n \sqrt{h}).
\end{multline}
Through the Fourier representation of the delta function, we conclude
\begin{multline}
P[x^{(N)},v^{(N)}\vert \omega^{(N}] =\\ \int\prod_{j=0}^{N-1}\frac{dk_j}{2\pi}\frac{dq_j}{2\pi} e^{-i\sum_j k_j\left(x_{j+1}-x_j-hv_j\right)}\\ 
\times e^{- i\sum_j q_j\left(v_{j+1} - v_j + \kappa v_j h - f_jh - \sqrt{C}\sqrt{h}\omega_j\right)}.
\end{multline}

We undo the conditioning by integrating in $\omega^{(N)}$ with Gaussian weight. We find
\begin{multline}
     P[x^{(N)},v^{(N)}] =\\ \int \prod_{j=0}^{N-1}\frac{dk_j}{2\pi}\frac{dq_j}{2\pi} e^{-i\sum_j k_j\left(x_{j+1}-x_j-hv_j\right)}\\\times e^{- i\sum_j q_j\left(v_{j+1} - v_j + \kappa v_j h - f_jh\right) + \sum_j C (iq_j)^2 h/2}.
\end{multline}
This should be compared with Eq. $\eqref{Discrete Path Integral}$, as it gives the path-integral expression in the formal limit $h\to 0$: we let $ik_j\to\tilde{x}$, $iq_j\to\tilde{v}$ and replace $\int \prod_{j=0}^{N-1}\frac{dk_j}{2\pi}\frac{dq_j}{2\pi}$ by the path integral measure $\mathcal{D}\tilde{x}\mathcal{D}\tilde{v}$. Finally, we arrive at Eq. $\eqref{Path integral expression for P[x,v]}$ for the probability density functional $P[x,v]$ and integrate out $v$ to get Eq. $\eqref{Path integral expression for P[x], second order case}$.

\subsection*{Appendix F: Simulations}

The simulations discussed in the main text were performed using the following discrete version of Eq. \ref{eq:SDE},
\begin{equation}
    x(t+\Delta t)=x(t)+f(x(t))\Delta t+\sqrt{D}\eta(t)
\end{equation}
where $\Delta t$ is the time interval between iterations. In our case, we chose this time interval to be $50\mu$s, which correspond to using a sampling frequency of $20$kHz. In the data displayed in Figs. \ref{fig:Figure2} and \ref{fig:Figure3}, each point in a PSD is the mean value between the corresponding values from 10 individual PSD's, while each errorbar is the standard deviation obtained from those 10 points. Each of these 10 PSDS's was obtained by averaging 500 PSD's calculated from simulations of 0.5 seconds of duration.

In Fig. \ref{fig:Figure3}, the final PSD's were fitted to a Lorentzian function,
\begin{equation}
    S(f)=\frac{A}{f_c^2+f^2}.
\end{equation}
In order to neglect the effects of aliasing \cite{BergSorensen2004},the fits were performed using only frequencies smaller than 1kHz, a value 10 times smaller than the Nyquist frequency.
To find the power law between the corner frequency and the parameters $a$ and $D$, the values of $f_c$ obtained while keeping $D$ constant were fitted to a function of the form $Ka^L$, and the ones obtained while keeping $a$ constant were fitted to $MD^N$.

\bibliographystyle{unsrt}
\bibliography{main.bib}

\end{document}